%% file: gassert-demo.tex
\PassOptionsToPackage{table}{xcolor}

\documentclass[conference]{IEEEtran}
\IEEEoverridecommandlockouts
\usepackage{cite}

\usepackage{amsmath,amssymb,amsfonts}
\usepackage{algorithmic}
\usepackage{graphicx}
\usepackage{textcomp}
\usepackage{xcolor}
 \usepackage[hidelinks]{hyperref}
\usepackage{amsmath}
\usepackage{algorithmic}
\usepackage{array}
\usepackage{fixltx2e}
\usepackage{stfloats}
\usepackage{url}
\usepackage{nicefrac}
\usepackage[american]{babel}
\usepackage[utf8x]{inputenc}
\usepackage[T1]{fontenc}
\usepackage{csquotes}
\usepackage[activate={true,nocompatibility},final,kerning=true,spacing=true,factor=1100,stretch=10,shrink=10]{microtype}

\usepackage[htt]{hyphenat}
\usepackage{paralist}
\usepackage{enumitem}
\usepackage{hyperref}
\usepackage{url}
\usepackage[all]{hypcap}
\usepackage{flushend}
\usepackage{etoolbox}

\usepackage{amsmath}
\usepackage{siunitx}
\usepackage{commath}
\robustify\bfseries

\newsavebox\CBox

\usepackage[super]{nth}

\usepackage{booktabs}
\usepackage{multirow}
\usepackage{color, colortbl}

\usepackage{makecell}

\usepackage{textcomp}
\usepackage{float}

\usepackage{graphicx}
\usepackage{tcolorbox}

\usepackage[ruled, vlined, linesnumbered]{algorithm2e}
\SetAlFnt{\small}
\SetAlCapNameFnt{\small}
\SetAlCapFnt{\small}
\SetFuncSty{textsc}
\AtBeginEnvironment{algorithm}{
	\DontPrintSemicolon
}
\usepackage{inconsolata}
\definecolor{pblue}{rgb}{0.13,0.13,1}
\definecolor{pgreen}{rgb}{0,0.5,0}
\definecolor{pred}{rgb}{0.9,0,0}
\definecolor{pgrey}{rgb}{0.46,0.45,0.48}
\usepackage{listingsutf8}
\lstnewenvironment{java}{\lstset{style=java}}{}

\usepackage{lipsum}
\usepackage{xparse}
\usepackage{xspace}

\RenewDocumentCommand{\paragraph}{m}{\smallskip\noindent \textbf{#1.}}

\newcommand{\oasis}{\textsc{OASIs}\@\xspace}
\newcommand{\tool}{\textsc{GAssert}\@\xspace}
\newcommand{\daikon}{\textsc{Daikon}\@\xspace}
\newcommand{\randoop}{\textsc{Randoop}\@\xspace}
\newcommand{\evosuite}{\textsc{EvoSuite}\@\xspace}
\newcommand{\random}{\textsc{Random}\@\xspace}
\newcommand{\inv}{\textsc{Inv-based}\@\xspace}

\newcommand{\major}{\textsc{Major}\@\xspace}
\newcommand{\pit}{\textsc{PIT}\@\xspace}
\newcommand{\Java}{\textsc{Java}\@\xspace}

\newtcolorbox{custombox}[1]{
	colback=gray!5,
	colframe=gray!70,
	boxrule=0.4mm,
	left=1.5mm,
	right=1.5mm,
	top=1.5mm,
	bottom=1.5mm,
	fonttitle=\bfseries,
	title=#1
}

\usepackage{pifont}
\usepackage{color}

\def\BibTeX{{\rm B\kern-.05em{\sc i\kern-.025em b}\kern-.08em
    T\kern-.1667em\lower.7ex\hbox{E}\kern-.125emX}}

\begin{document}

\title{GAssert: A Fully Automated Tool \\to Improve Assertion Oracles}

\author{
	\IEEEauthorblockN{Valerio Terragni\IEEEauthorrefmark{1}, Gunel Jahangirova\IEEEauthorrefmark{2}, Paolo Tonella\IEEEauthorrefmark{2} and Mauro Pezzè\IEEEauthorrefmark{2}\IEEEauthorrefmark{3}}
		\IEEEauthorblockA{\IEEEauthorrefmark{1}University of Auckland, Auckland, New Zealand}
	\IEEEauthorblockA{\IEEEauthorrefmark{2}Università della Svizzera italiana, Lugano, Switzerland}
		\IEEEauthorblockA{\IEEEauthorrefmark{3}Schaffhausen Institute of Technology, Schaffhausen, Switzerland \\
		v.terragni@auckland.ac.nz \{gunel.jahangirova, paolo.tonella, mauro.pezze\}@usi.ch } 	\vspace{-2mm} \\
\vspace{-2mm}
	\emph{This is the author's version of the work. The definitive version appeared at ICSE 2021} 
}

\maketitle

\begin{abstract}
This demo presents the implementation and usage details of \tool, the first tool to automatically improve assertion oracles.
Assertion oracles are executable boolean expressions placed inside the program that should pass (return true) for all correct executions and fail (return false) for all incorrect executions. 
Because designing perfect assertion oracles is difficult, assertions are prone to both false positives (the assertion fails but should pass) and false negatives (the assertion passes but should fail).
Given a Java method containing an assertion oracle to improve, \tool returns an improved assertion with fewer false positives and false negatives than the initial assertion.
Internally, \tool implements a novel co-evolutionary algorithm that explores the space of possible assertions guided by two fitness functions that reward assertions with fewer false positives, fewer false negatives, and smaller size.
\end{abstract}


\section{Introduction}

Assertion oracles (also called program assertions) are exe\-cutable boolean expressions  placed inside the program that predicate on the values of variables at specific program points.
A perfect assertion oracle passes (returns true) for all correct executions and fails (returns false) for all incorrect executions~\cite{barr:survey:tse:2015}.
For most non-trivial programs, designing perfect oracles is difficult, and thus assertion oracles often fail to distinguish between correct and incorrect executions~\cite{jahangirova:oracle:issta:2016}, that is, they are prone to both false positives and false negatives.
A \textit{false positive} is a correct program state in which the assertion fails (but should pass). A \textit{false negative} is an incorrect program state in which the assertion passes (but should fail).
False positives and false negatives are jointly called \emph{oracle deficiencies}~\cite{jahangirova:oracle:issta:2016}.

Improving the quality of assertion oracles by removing their deficiencies is an important problem and would bring several benefits, primarily the reduced false alarm rate and increased fault detection capability of test suites.
Notably, automated test case generators~\cite{Pacheco:Randoop:ICSE:2007,Fraser:Evosuite:TSE:2013} will benefit the most from improved assertion oracles.
This is because high quality assertion oracles  avoid  the need to automatically define a test oracle for each generated test case. 
Indeed, the \emph{oracle problem}~\cite{barr:survey:tse:2015} is a major obstacle in test automation, limiting the effectiveness of automatically generated test suites~\cite{Pacheco:Randoop:ICSE:2007,Fraser:Evosuite:TSE:2013}.

\smallskip
Recently, Jahangirova et al. proposed the \oasis~\cite{jahangirova:oracle:issta:2016} approach to automatically generate test cases and mutations that expose the oracle deficiencies of a given assertion oracle.
The \oasis's output is intended to help developers improve the oracles.
However, even with the oracle deficiencies provided by \oasis, manually improving assertion oracles  remains difficult~\cite{Jahangirova:tse:2019}. 
In fact, the authors of \oasis report that for only 67\% of the given assertions humans successfully removed all oracle deficiencies reported by \oasis~\cite{Jahangirova:tse:2019}.

\textbf{\tool}~\cite{Terragni:gassert:ESECFSE:2020} (\emph{\textbf{\underline{G}enetic \underline{ASSERT}ion improvement}}) is the first technique to automatically improve assertion oracles. 
The envisioned users of \tool are \Java developers who wish to improve the quality of their assertion oracles.
Given an assertion oracle $\alpha$ and its oracle deficiencies provided by \oasis, \tool explores the space of possible assertions to return a new assertion $\alpha^\prime$ with zero false positives and the smallest number of false negatives.
\tool favors assertions with zero false positives, as false alarms trigger an expensive debugging process.

Internally, \tool implements a novel  co-evolutionary algorithm  that evolves two populations of assertions in parallel with three competing objectives:
\begin{inparaenum}[(i)]
	\item minimizing the number of false positives,
	\item minimizing the number of false negatives,
	\item minimizing the size of the assertion.
\end{inparaenum}
The first population rewards solutions with fewer false positives, the second population those with fewer false negatives, considering the remaining objectives only in tie cases.
On a regular basis, the two populations exchange their best individuals (population migration) to supply the other population with good genetic material useful to improve the secondary objective.

We evaluated the ability of \tool to improve an initial set of \daikon~\cite{Ernst:daikon:icse:1999} generated assertions on 34 methods from 7 \Java code bases~\cite{Terragni:gassert:ESECFSE:2020}.
The  improved assertions have  zero false positives and reduce the false negatives by 40\% (on average) with respect to the initial \daikon assertions. 
When evaluated with unseen tests and mutants, the assertions generated by \tool increase the mutation score by 34\% (on average).

This paper extends our recent ESEC/FSE paper~\cite{Terragni:gassert:ESECFSE:2020} by giving details about the design, implementation and usage of \tool, which can be found at:
\vspace{-1mm}
\begin{center}
	 \texttt{\small \url{https://github.com/valerio-terragni/gassert}}
\end{center}

\section{\tool}

\begin{figure*}[th]
	\centering

	\resizebox{1\textwidth}{!}{%
		\includegraphics{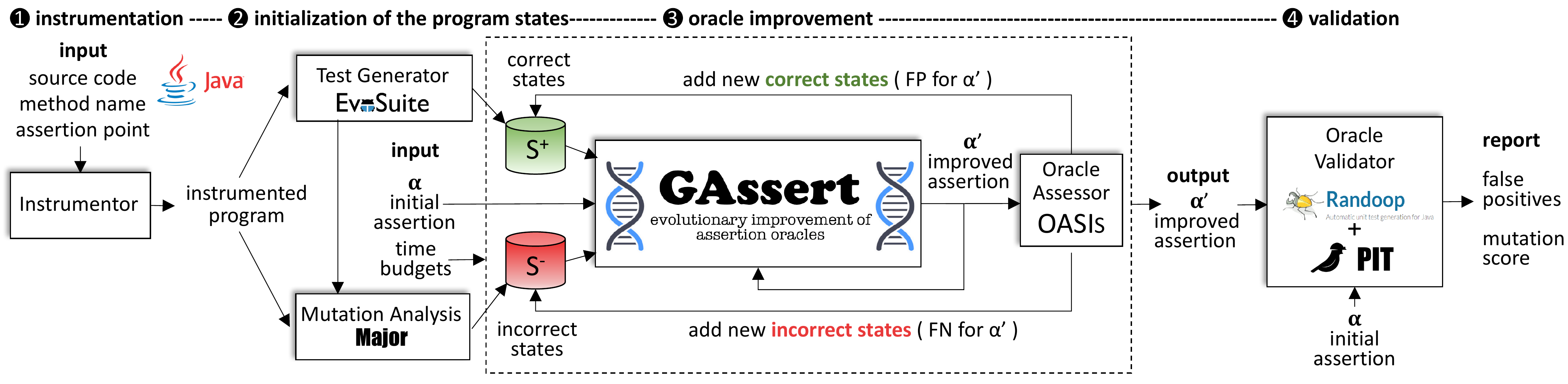}
	}
	\vspace{-8mm}
	\caption{Logical workflow of \tool}
		\vspace{-2mm}
	\label{fig:overview}
\end{figure*}

\tool is a command-line tool, which has five inputs:
\begin{inparaenum}[(i)]
	\item the source code  of a \Java class,
	\item the name of the method under analysis,
	\item an initial assertion oracle $\alpha$,
	\item the point in the method were  $\alpha$ is placed (assertion point), and 
	\item a global and an internal  time budget for the oracle improvement.
	\end{inparaenum}
The output of \tool is an improved assertion oracle $\alpha^\prime$.
Figure~\ref{fig:overview} shows the logical workflow of \tool, which is composed of four phases:

\ding{182} \textbf{instrumentation}, \tool instruments the method under analysis to capture program states at runtime.
 
\ding{183} \textbf{initialization of the program states}, \tool produces an initial set of correct and incorrect states ($\mathcal{S}^\text{+}$ and $\mathcal{S}^\text{-}$) by executing an initial test suite on the instrumented version of the original method and on its faulty variants (mutations). 

 \ding{184} \textbf{oracle improvement}, this phase alternatively executes \tool and \oasis until  a time budget expires or \oasis does not find oracle deficiencies for the improved oracle $\alpha^\prime$.

\ding{185} \textbf{validation}, \tool evaluates the initial and improved  assertions ($\alpha$ and $\alpha^\prime$) on a validation set of tests and mutations.

\smallskip
Figure~\ref{fig:examples} shows a running example of \tool applied to  method \texttt{\small floor} of  class \texttt{\small FastMath} of \texttt{\small The Apache Commons Math library}. 
The method implements a fast algorithm for the floor computation, which takes as input a real number \texttt{\small x}, and outputs the greatest integer less than or equal to \texttt{\small x} (\texttt{\small result}).

Figure~\ref{fig:examples} (a) shows the instrumented version of the method and the initial assertion $\alpha$ : \texttt{\small(y == result) \&\& (x > result)} (line~16).
Although  $\alpha$ properly behaves in many correct and incorrect states, it has both false positives and false negatives.

Figure~\ref{fig:examples} (b) shows an example of a false positive program state for $\alpha$,  together with the \evosuite test case that produces it.
For this correct state, \texttt{\small x} is not greater than \texttt{\small result}, and thus  $\alpha$ returns false (but should return true).

Figure~\ref{fig:examples} (c) shows an example of a false negative program state for $\alpha$,  together with the \evosuite test case and \major mutation that produce it.
For this incorrect state, the values of \texttt{\small y} and \texttt{\small result} are wrong (they should be \texttt{\small 0}), but $\alpha$ does not fail (it returns true instead of false).

Running \tool provides an improved assertion $\alpha^\prime$: \\ \texttt{\small(y == result) \&\& (x >= result) \&\& (x < (result+1))}, which intuitively captures the 
behavior of a \emph{floor} function, and it does not suffer from the  oracle deficiencies of Figure~\ref{fig:examples} (b) and (c).
The following subsections describe the four phases in detail.

\subsection{Instrumentation}

The \emph{Instrumentor} component of \tool inserts additional method calls in the method under analysis to collect program states at runtime.
We implemented it by relying on the source code manipulator  \textsc{JavaParser} (v. 3.6.26)\footnote{\url{http://javaparser.org/}}.

The instrumentor analyzes  the source code of the method under analysis to collect the method parameters (including the object receiver) \textit{MP}, and all the local variables \textit{LV} that  are visible at the assertion point. 
For the example in Figure~\ref{fig:examples}, \textit{MP} $= \{\texttt{\small x} \}$ and \textit{LV} $= \{\texttt{\small y}, \texttt{\small result}\}$.
It then instruments two method calls.
The first one is placed at the beginning of the method passing \textit{MP} as an argument (line~3 in Figure~\ref{fig:examples}).
The second one is located right before the assertion point passing both \textit{MP} and \textit{LV} as arguments (line~15 in Figure~\ref{fig:examples}).
By considering the parameter values at the beginning of the method, \tool is able to generate assertions that predicate on method preconditions.
\tool distinguishes the variables in \textit{MP}  at the two execution points by adding the prefix \enquote{\texttt{\small old\_}} to the variable names at the first execution point.

\subsection{Initialization of the Program States }

\begin{figure*}[th]
	\centering
	\vspace{-2mm}
	\resizebox{1\textwidth}{!}{%
		\includegraphics{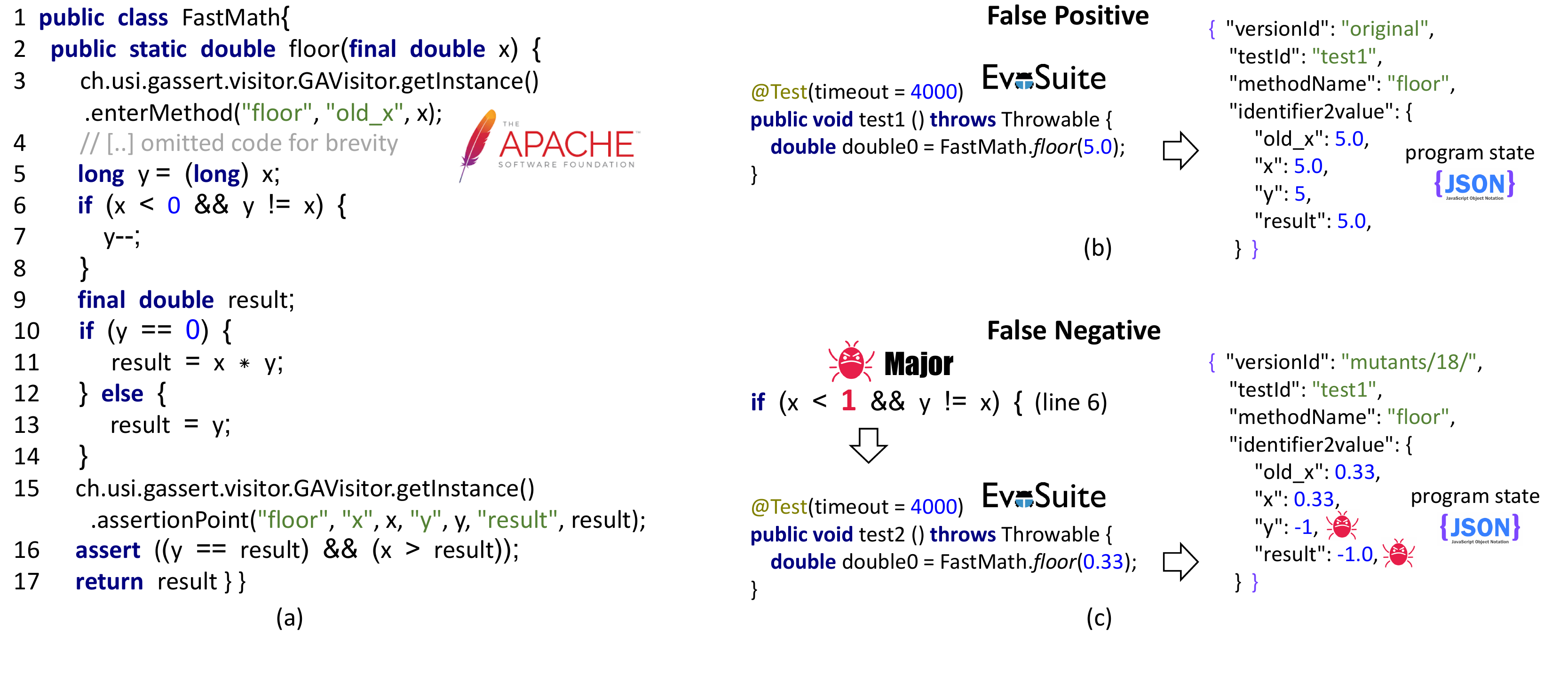}
	}
	\vspace{-15mm}
	\caption{Instrumented version of the method floor (a), examples of a false positive (b) and a false negative (c) for the assertion at line 17.}
			\vspace{-2mm}
	\label{fig:examples}
\end{figure*}

\tool  needs test cases and mutations to initialize the repositories of correct and incorrect states ($\mathcal{S}^\text{+}$ and $\mathcal{S}^\text{-}$).
Such repositories are progressively filled with the correct and incorrect states returned by \oasis.
The rationale for initializing the repositories, instead of  immediately relying on states returned by \oasis,  is to minimize the number of iterations.
In this way, we avoid using \oasis to detect obvious oracle deficiencies, and rather let \oasis focus on hard-to-find ones.

To enable full automation, \tool obtains the initial test cases and mutations from \evosuite~\cite{Fraser:Evosuite:TSE:2013} (v. 1.0.6)\footnote{\url{https://www.evosuite.org/}} and \major(v. 1.3.4)\footnote{\url{https://mutation-testing.org/}}, respectively.
\evosuite is a search-based test case generator for \Java driven by various coverage criteria. In our experiments we used the branch coverage criterion. 
\major generates source-code mutants of a \Java class by seeding artificial faults. 
Notably, developers and testers can also provide additional manually-written test cases and  faulty versions of the method under analysis, which likely yield to assertion oracles of higher quality.
Figure~\ref{fig:examples} (b) and (c) show examples of test cases and mutations generated at phase \ding{183}.

\tool initializes $\mathcal{S}^\text{+}$ by executing the test cases on the instrumented method, and $\mathcal{S}^\text{-}$ by executing the test cases on the instrumented mutations.
Invoking the instrumented code constructs the program states and saves them on disk as JSON files.
Saving the states is crucial  to quickly calculate how many false positives and false negatives a candidate assertion has, without requiring  expensive program re-executions.
Indeed, our evolutionary algorithm might explore millions of candidate assertions before converging to an optimized solution.

\tool generates assertion oracles as Boolean expressions that predicate on variables and functions of Boolean or numerical types (see Table~\ref{table:functions}).
As such,  a program state $s$ is a set of Boolean or numerical variables $\{ v_1, \cdots , v_k \}$.
Each variable $v_i$ has a type, an identifier, and a value. 
For each variable $v_i$ passed as arguments to the instrumented method calls, \tool constructs a program state $s$ as follows:

If~$v_i$ is of primitive type (Boolean or numerical), \tool simply adds its runtime value to $s$ (rounding floats with a fixed precision) using the variable name as  identifier.
If~$v_i$ is non-primitive (objects in \Java), \tool needs to convert $v_i$ into primitive values.
\tool achieves this with a \emph{hybrid state serialization} that combines the object serialization and observer abstraction approaches~\cite{Terragni:gassert:ESECFSE:2020}.
\emph{Object serialization} captures the values of primitive-type object fields that are recursively reachable from $v_i$.
\emph{Observers abstraction} captures the return values of observer methods invoked with $v_i$ as the object receiver.
\emph{Observer methods} are side-effect free
methods that are declared in $v_i$'s class and return primitive values.

\tool automatically finds the observer methods of $v_i$'s~class with an efficient (but conservative) byte-code static analyzer.
The analyzer marks a method as side-effect free if it cannot  directly or indirectly execute \texttt{putfield} or \texttt{putstatic} bytecode instructions.
\tool relies on \Java \textsc{Reflection} to get at runtime the values of primitive fields and the return values of the observer methods invocations.
Figure~\ref{fig:examples} (b) and~(c) show the JSON files of the program states obtained when executing the corresponding test cases. 


\input{tables/table-functions}

\subsection{Oracle Improvement}
The oracle improvement process takes in input an initial assertion $\alpha$ and two time budgets (an internal one for the \mbox{co-evolutionary} algorithm and a global one for the whole process), and outputs an improved assertion $\alpha^\prime$.
The initial assertion can be specified by the user or automatically generated by our scripts using the invariant generator \daikon~(v. 5.7.2)
~\cite{Ernst:daikon:icse:1999}.
The default configuration is an internal time budget of 30 minutes and a global one of 90 minutes.

The oracle improvement process is composed of three steps:

\noindent \textbf{I.} \tool executes the co-evolutionary algorithm and terminates when it finds an assertion $\alpha^\prime$ with zero oracle deficiencies with respect to the current correct and incorrect states ($\mathcal{S}^\text{+}$~and~$\mathcal{S}^\text{-}$)  or the internal time budget expires.
In the latter case, \tool returns the assertion $\alpha^\prime$ that among all the explored assertions with zero false positives has  the lowest number of false negatives.

\noindent \textbf{II.} \oasis searches for oracle deficiencies of $\alpha^\prime$. If it finds them, it adds the resulting program states to $\mathcal{S}^\text{+}$ and $\mathcal{S}^\text{-}$.

\noindent \textbf{III.} \tool takes in input $\alpha^\prime$ and repeats step  I.

Such an iterative process terminates when \oasis does not find oracle deficiencies for $\alpha^\prime$  or the global time budget expires.

The co-evolutionary algorithm evolves two populations of assertions in parallel. One population uses the number of false positives as the fitness score, while the other uses the number of false negatives.
The remaining objectives are used only in tie cases. 
The two populations periodically exchange their best individuals to help optimize  the secondary objectives.

\tool evolves each population following the GP approach:
\begin{inparaenum}[(i)]
	\item \textbf{selection}: selecting pairs of assertions (parents) by means of fitness functions that reward fitter solutions;
	\item \textbf{crossover}: creating new (and possibly fitter) offspring by combining portions of the parent assertions; and,
	\item \textbf{mutation}: mutating the offspring (with a certain probability).
\end{inparaenum}
\tool adopts a tree-like representation of assertions and uses the standard  tree-based mutation and crossover operators. Moreover, 
we propose novel selection and crossover operators that are specific to the oracle improvement problem~\cite{Terragni:gassert:ESECFSE:2020}.

\smallskip
We now exemplify how our  algorithm could obtain $\alpha^\prime$ : \\ \texttt{\small(y == result) \&\& (x >= result) \&\& (x < (result+1))} in the example of Figure~\ref{fig:examples}.
Let us assume that the algorithm selects two parents $\alpha_{p1}$ :  \texttt{\small(y == result) \&\& (x > result)} and $\alpha_{p2}$ : \texttt{\small (x < (result+1))}.
If  crossover produced  \texttt{\small(y == result) \&\& (x > result) \&\& (x < (result+1))} and  mutation changed \texttt{\small >} to \texttt{\small >=}, \tool would obtain the improved assertion $\alpha^\prime$.

\oasis~\cite{jahangirova:oracle:issta:2016} detects false positives of an assertion $\alpha$ by  creating a new branch with the negated boolean expression of $\alpha$.
It then uses search-based test generation~\cite{Fraser:Evosuite:TSE:2013} to produce test cases that cover the branch and consequently make the assertion fail. 
For instance, given the assertion at line~17 in Figure~\ref{fig:examples} (a), \oasis would  obtain the test case in Figure~\ref{fig:examples} (b) by replacing the assertion with the artificial branch \texttt{\small if((y != result) || (x <= result))\{\}} and then driving search-based test generation towards covering this branch.

\oasis detects  false negatives by combining test case generation and mutation testing.
It injects faults into the program and generates test cases for the faulty version such that at least one of the variables used in the assertion changes its value, while the outcome of the assertion does not change. 

\subsection{Validation}

To evaluate if the improved assertions generalize well with unseen correct and incorrect states, we generated new test cases and mutations using \randoop~\cite{Pacheco:Randoop:ICSE:2007} (v. 4.2.0)\footnote{\url{https://randoop.github.io/randoop/}} and \pit (v. 1.4.0)\footnote{\url{https://pitest.org/}}, respectively.
Because they are different tools from the ones that provide test cases and mutations to the oracle improvement process (\evosuite, \major and \oasis), they are expected to provide different test cases and mutations.

The validation phase counts the number of validation tests that fail with
 the improved assertion inserted at the assertion point.
If it is zero, we use \pit to run the validation tests and report the mutation score. 
If it is greater than zero, we cannot run \pit  because we need a green test suite. 
In such case, if the evaluated assertion has the form \texttt{\small assert($\alpha_1$ \texttt{\small{\&\&}} $\alpha_2$ \texttt{\small{\&\&}} $\alpha_3$)}, \tool considers each of the smaller assertions  \texttt{\small assert($\alpha_1$)}, \texttt{\small assert($\alpha_2$)}  and \texttt{\small assert($\alpha_3$)} removing those that have false positives.
Then, it concatenates the remaining smaller conditions with \texttt{\small{\&\&}} and it performs mutation testing with \pit for this reduced assertion. 
It then repeats this process for the initial assertion.
The user of \tool can compare the HTML reports of PIT to better understand the fault detection capability of the initial and improved assertion oracles.



\section{Evaluation}
We evaluated \tool on 34 methods from 7 different Java code bases~\cite{Terragni:gassert:ESECFSE:2020}.
The validation phase shows that improved assertions  have always zero false positives and achieve, on average, 34\% increase in mutation score with respect to initial \daikon assertions. 

In addition, we compared \tool to two baselines \random and \inv.
\random is a variant of \tool with no guidance by the fitness functions.
The results show that \tool-generated assertions  outperform the \random-generated ones for 50\% of the subjects.
\inv executes \daikon on the initial test suite and obtains an invariant. 
It then augments the initial test suite with the test cases generated by \oasis that reveal false positives, and re-executes \daikon. 
This process repeats until 
the global time budget expires. 
\tool outperforms \inv  for 63\% of the subjects.

We also compared \tool with a set of human-improved assertions collected from 102 developers
~\cite{Jahangirova:tse:2019}. 
Our results show that the  mutation score achieved by \tool is always equal to the average mutation score achieved by the humans. 
Moreover, 10\% of the human-improved assertions achieve a lower mutation score than the assertions improved by \tool.

\section{Conclusion}

While there are many techniques to automatically generate program assertions (e.g., program invariants~\cite{Ernst:daikon:icse:1999}), automatically improving assertion oracles by removing false positives and false negatives is an unexplored problem.
\tool is the first technique of its kind,  opening a new research area.

Techniques like \tool might encourage developers to use assertion oracles  more often,  resulting in better software quality in the long run. 
We highlight three promising future research directions:
\begin{inparaenum}[(i)]
\item leverage the feedback of \oasis not only after \tool has produced the final assertion, but also during the evolution,
\item increase the expressiveness of the assertions (e.g., with universal quantifiers), and 
\item make the improved assertions easier to  read and understand for humans.
\end{inparaenum}

\bibliographystyle{IEEEtran}

\end{document}

%% file: tables/table-functions.tex

\begin{table}[t]
\centering
\renewcommand*{\arraystretch}{1}
\setlength{\tabcolsep}{4pt}
\vspace{-3mm}
\caption{Functions considered by \tool}
\vspace{-3mm}
\label{table:functions}
	\resizebox{\linewidth}{!}{%
				\rowcolors{1}{}{gray!10}
\begin{tabular}{lll}
				\hiderowcolors

	\toprule
	\textbf{operand} & \textbf{output} & \textbf{functions} \vspace{-1mm} \\ 
		\textbf{type} & \textbf{type}  \\
	\midrule
\showrowcolors
$\langle$ number, number $\rangle$	&    number             &  +, *, -, /, \% (modulo)                 \\
$\langle$ number, number $\rangle$	&    boolean             &  ==, <, >, $\leq$, $\geq$, $\neq$                \\
$\langle$ boolean, boolean $\rangle$	&    boolean             &  AND, OR, XOR, EXOR,  $\rightarrow$ (implies), == (equiv.)    \\
$\langle$ boolean $\rangle$	&    boolean             &    NOT               \\
	\bottomrule
\end{tabular}
}
\end{table}